\documentclass[twocolumn,5p,times,aps,prl,superscriptaddress,showkeys]{revtex4-1}
\usepackage{graphicx}
\usepackage{amssymb}
\usepackage[amssymb]{SIunits}
\usepackage{subfigure}
\usepackage{float}
\usepackage{amsmath}
\usepackage{color}
\usepackage{ulem}
\usepackage{multirow}
\usepackage{wasysym}
\newcommand{\lamo}{$\Lambda$}
\newcommand{\lam}{\lamo\ }
\newcommand{\sigo}{$\Sigma$}

\newcommand{\Kpo}{K$^{+}$}

\newcommand{\nstaro}{N$^{*}$}
\newcommand{\nstar}{\nstaro\ }
\newcommand{\pKLo}{p\Kpo\lamo}

\newcommand{\pklo}{\pKLo}
\newcommand{\pKL}{\pKLo\ }
\newcommand{\pkL}{\pKLo\ }
\newcommand{\pkl}{\pKLo\ }
\newcommand{\distoo}{DISTO}

\newcommand{\disto}{\distoo\ }

\newcommand{\DISTO}{\disto\ }
\newcommand{\cosyo}{COSY-TOF}

\newcommand{\cosy}{\cosyo\ }

\newcommand{\COSY}{\cosy}

\newcommand{\fig}[1]{Figure~\ref{#1}}   
\newcommand{\tab}[1]{Table~\ref{#1}}

\newcommand{\Fig}[1]{Figure~\ref{#1}}
\newcommand{\Tab}[1]{Table~\ref{#1}}

\begin{document}
\setlength{\unitlength}{0.02\columnwidth}
\title{Determination of N* amplitudes from associated strangeness production in p+p collisions} 
\author{R. M\"unzer}\email{ robert.muenzer@cern.ch}\affiliation{Excellence Cluster Universe, Technische Universit\"at M\"unchen,Boltzmannstr. 2, D-85748, Germany}\affiliation{Physik Department E62, Technische Universit\"at M\"unchen, 85748~Garching, Germany}
\author{L. Fabbietti}\email{ laura.fabbietti@ph.tum.de}\affiliation{Excellence Cluster Universe, Technische Universit\"at M\"unchen,Boltzmannstr. 2, D-85748, Germany}\affiliation{Physik Department E62, Technische Universit\"at M\"unchen, 85748~Garching, Germany}
\author{E. Epple}\affiliation{Yale University, New Haven, Connecticut, United States}
\author{S. Lu}\affiliation{Physik Department E62, Technische Universit\"at M\"unchen, 85748~Garching, Germany}
\author{P. Klose}\affiliation{Physik Department E62, Technische Universit\"at M\"unchen, 85748~Garching, Germany}
\author{F. Hauenstein}\affiliation{Institut f\"ur Kernphysik, Forschungszentrum J\"ulich, 52428 J\"ulich, Germany}
\author{N. Herrmann}\affiliation{Physikalisches Institut der Universit\"{a}t Heidelberg, Heidelberg, Germany}
\author{D. Grzonka}\affiliation{Institut f\"ur Kernphysik, Forschungszentrum J\"ulich, 52428 J\"ulich, Germany}\affiliation{J\"ulich Aachen Research Alliance, Forces and Matter Experiments (JARA-FAME) }\affiliation{Experimentalphysik I, Ruhr-Universit\"at Bochum, 44780 Bochum, Germany}
\author{Y. Leifels}\affiliation{GSI Helmholtzzentrum f\"ur Schwerionenforschung GmbH, 64291 Darmstadt, Germany}
\author{M. Maggiora}\affiliation{Istituto Nazionale di Fisica Nucleare (INFN) - Sezione di Torino, 10125 Torino, Italy}
\author{D. Pleiner}\affiliation{Physik Department E62, Technische Universit\"at M\"unchen, 85748~Garching, Germany}
\author{B. Ramstein}\affiliation{Institut de Physique Nucleaire, CNRS/IN2P3 - Univ. Paris Sud, F-91406 Orsay Cedex, France}
\author{J. Ritman}\affiliation{Institut f\"ur Kernphysik, Forschungszentrum J\"ulich, 52428 J\"ulich, Germany}\affiliation{J\"ulich Aachen Research Alliance, Forces and Matter Experiments (JARA-FAME) }\affiliation{Experimentalphysik I, Ruhr-Universit\"at Bochum, 44780 Bochum, Germany}
\author{E. Roderburg}\affiliation{Institut f\"ur Kernphysik, Forschungszentrum J\"ulich, 52428 J\"ulich, Germany}
\author{P. Salabura}\affiliation{Smoluchowski Institute of Physics, Jagiellonian University of Cracow, 30-059~Krak\'{o}w, Poland}
\author{A.Sarantsev}\affiliation{Petersburg Nuclear Physics Institute, Gatchina, Russia}
\author{Z.~Basrak}\affiliation{Ru{d\llap{\raise 1.22ex\hbox{\vrule height 0.09ex width 0.2em}}\rlap{\raise 1.22ex\hbox{\vrule height 0.09ex width 0.06em}}}er Bo\v{s}kovi\'{c} Institute, Zagreb, Croatia}
\author{P.~Buehler}\affiliation{Stefan-Meyer-Institut f\"{u}r subatomare Physik, \"{O}sterreichische Akademie der Wissenschaften, Wien, Austria}
\author{M.~Cargnelli}\affiliation{Stefan-Meyer-Institut f\"{u}r subatomare Physik, \"{O}sterreichische Akademie der Wissenschaften, Wien, Austria}
\author{R.~\v{C}aplar}\affiliation{Ru{d\llap{\raise 1.22ex\hbox{\vrule height 0.09ex width 0.2em}}\rlap{\raise 1.22ex\hbox{\vrule height 0.09ex width 0.06em}}}er Bo\v{s}kovi\'{c} Institute, Zagreb, Croatia}
\author{H.~Clement}\affiliation {Physikalisches Institut der Universit\"at T\"ubingen, Auf der Morgenstelle 14, 72076 T\"ubingen, Germany}\affiliation {Kepler Center for Astro and Particle Physics, University of T\"ubingen, Auf der Morgenstelle 14, 72076 T\"ubingen, Germany}
\author{O.~Czerwiakowa}\affiliation{Institute of Experimental Physics, Faculty of Physics, University of Warsaw, Warsaw, Poland}
\author{I.~Deppner}\affiliation{Physikalisches Institut der Universit\"{a}t Heidelberg, Heidelberg, Germany}
\author{M.~D\v{z}elalija}\affiliation{Faculty of Science, University of Split, Split, Croatia}
\author{W.~Eyrich}\affiliation{Friedrich-Alexander-Universit\"at Erlangen-N\"urnberg, 91058 Erlangen, Germany}
\author{Z.~Fodor}\affiliation{Wigner RCP, RMKI, Budapest, Hungary}
\author{P.~Gasik}\affiliation{Excellence Cluster Universe, Technische Universit\"at M\"unchen,Boltzmannstr. 2, D-85748, Germany}\affiliation{Physik Department E62, Technische Universit\"at M\"unchen, 85748~Garching, Germany}
\author{I.~Ga\v{s}pari\'c}\affiliation{Ru{d\llap{\raise 1.22ex\hbox{\vrule height 0.09ex width 0.2em}}\rlap{\raise 1.22ex\hbox{\vrule height 0.09ex width 0.06em}}}er Bo\v{s}kovi\'{c} Institute, Zagreb, Croatia}
\author{A.~Gillitzer}\affiliation{Institut f\"ur Kernphysik, Forschungszentrum J\"ulich, 52428 J\"ulich, Germany}\affiliation{J\"ulich Aachen Research Alliance, Forces and Matter Experiments (JARA-FAME) }\affiliation{Experimentalphysik I, Ruhr-Universit\"at Bochum, 44780 Bochum, Germany}
\author{Y.~Grishkin}\affiliation{Institute for Theoretical and Experimental Physics, Moscow, Russia}
\author{O.N.~Hartmann}\affiliation{GSI Helmholtzzentrum f\"ur Schwerionenforschung GmbH, 64291 Darmstadt, Germany}
\author{K.D.~Hildenbrand}\affiliation{GSI Helmholtzzentrum f\"ur Schwerionenforschung GmbH, 64291 Darmstadt, Germany}
\author{B.~Hong}\affiliation{Korea University, Seoul, Korea}
\author{T.I.~Kang}\affiliation{GSI Helmholtzzentrum f\"ur Schwerionenforschung GmbH, 64291 Darmstadt, Germany} \affiliation{Korea University, Seoul, Korea}
\author{J.~Kecskemeti}\affiliation{Wigner RCP, RMKI, Budapest, Hungary}
\author{Y.J.~Kim}\affiliation{GSI Helmholtzzentrum f\"ur Schwerionenforschung GmbH, 64291 Darmstadt, Germany}
\author{M.~Kirejczyk}\affiliation{Institute of Experimental Physics, Faculty of Physics, University of Warsaw, Warsaw, Poland}
\author{M.~Ki\v{s}}\affiliation{GSI Helmholtzzentrum f\"ur Schwerionenforschung GmbH, 64291 Darmstadt, Germany}
\author{P.~Koczon}\affiliation{GSI Helmholtzzentrum f\"ur Schwerionenforschung GmbH, 64291 Darmstadt, Germany}
\author{R.~Kotte}\affiliation{Institut f\"{u}r Strahlenphysik, Helmholtz-Zentrum Dresden-Rossendorf, Dresden, Germany}
\author{A.~Lebedev}\affiliation{Institute for Theoretical and Experimental Physics, Moscow, Russia}
\author{A.~Le F\`{e}vre}\affiliation{GSI Helmholtzzentrum f\"ur Schwerionenforschung GmbH, 64291 Darmstadt, Germany}
\author{J.L.~Liu}\affiliation{Harbin Institute of Technology, Harbin, China}
\author{V.~Manko}\affiliation{National Research Centre 'Kurchatov Institute', Moscow, Russia}
\author{J.~Marton}\affiliation{Stefan-Meyer-Institut f\"{u}r subatomare Physik, \"{O}sterreichische Akademie der Wissenschaften, Wien, Austria}
\author{T.~Matulewicz}\affiliation{Institute of Experimental Physics, Faculty of Physics, University of Warsaw, Warsaw, Poland}
\author{K.~Piasecki}\affiliation{Institute of Experimental Physics, Faculty of Physics, University of Warsaw, Warsaw, Poland}
\author{F.~Rami}\affiliation{Institut Pluridisciplinaire Hubert Curien and Universit\'{e} de Strasbourg, Strasbourg, France}
\author{A.~Reischl}\affiliation{Physikalisches Institut der Universit\"{a}t Heidelberg, Heidelberg, Germany}
\author{M.S.~Ryu}\affiliation{Korea University, Seoul, Korea}
\author{P.~Schmidt}\affiliation{Stefan-Meyer-Institut f\"{u}r subatomare Physik, \"{O}sterreichische Akademie der Wissenschaften, Wien, Austria}
\author{Z.~Seres}\affiliation{Wigner RCP, RMKI, Budapest, Hungary}
\author{B.~Sikora}\affiliation{Institute of Experimental Physics, Faculty of Physics, University of Warsaw, Warsaw, Poland}
\author{K.S.~Sim}\affiliation{Korea University, Seoul, Korea}
\author{K.~Siwek-Wilczy\'{n}ska}\affiliation{Institute of Experimental Physics, Faculty of Physics, University of Warsaw, Warsaw, Poland}
\author{V.~Smolyankin}\affiliation{Institute for Theoretical and Experimental Physics, Moscow, Russia}
\author{K.~Suzuki}\affiliation{Stefan-Meyer-Institut f\"{u}r subatomare Physik, \"{O}sterreichische Akademie der Wissenschaften, Wien, Austria}
\author{Z.~Tymi\'{n}ski}\affiliation{Physikalisches Institut der Universit\"{a}t Heidelberg, Heidelberg, Germany}
\author{P.~Wagner}\affiliation{Institut Pluridisciplinaire Hubert Curien and Universit\'{e} de Strasbourg, Strasbourg, France}
\author{I.~Weber}\affiliation{Faculty of Science, University of Split, Split, Croatia}
\author{E.~Widmann}\affiliation{Stefan-Meyer-Institut f\"{u}r subatomare Physik, \"{O}sterreichische Akademie der Wissenschaften, Wien, Austria}
\author{K.~Wi\'{s}niewski}\affiliation{Institute of Experimental Physics, Faculty of Physics, University of Warsaw, Warsaw, Poland}
\author{Z.G.~Xiao}\affiliation{Department of Physics, Tsinghua University, Beijing, China}
\author{T. Yamasaki}\affiliation{Department of Physics, The University of Tokyo, Tokyo, 113-0033}\affiliation{RIKEN Nishina Center, RIKEN, Wako, 351-0198, Japan}
\author{I.~Yushmanov}\affiliation{National Research Centre 'Kurchatov Institute', Moscow, Russia}
\author{P. Wintz}\affiliation{Institut f\"ur Kernphysik, Forschungszentrum J\"ulich, 52428 J\"ulich, Germany}
\author{Y.~Zhang}\affiliation{Institute of Modern Physics, Chinese Academy of Sciences, Lanzhou, China}
\author{A.~Zhilin}\affiliation{Institute for Theoretical and Experimental Physics, Moscow, Russia}
\author{V.~Zinyuk}\affiliation{Physikalisches Institut der Universit\"{a}t Heidelberg, Heidelberg, Germany}
\author{J.~Zmeskal}\affiliation{Stefan-Meyer-Institut f\"{u}r subatomare Physik, \"{O}sterreichische Akademie der Wissenschaften, Wien, Austria}
\graphicspath{{../figures/}}
\begin{abstract}
We present the first determination of the energy-dependent amplitudes of N$^{*}$ resonances extracted from their decay in
K$\Lambda$ pairs in p+p$\rightarrow$\pkl reactions.
A combined Partial Wave Analysis of seven data samples with exclusively reconstructed  p+p$\rightarrow$
\pkl events measured by the COSY-TOF, DISTO, FOPI and HADES Collaborations in fixed target experiments at kinetic energies 
between 2.14 to 3.5 GeV 
is used to determine the amplitude of the resonant and non-resonant contributions into the associated strangeness final state.
The contribution of seven N$^{*}$ resonances with masses between 1650 MeV/c$^{2}$ and 1900 MeV/c$^{2}$ for an excess
 energy between $0$ and $600$ MeV has been considered.
The \sigo-p cusp and final state interactions for the p-\lam channel are also included as coherent contributions in the PWA.
The N$^{*}$ contribution is found to be dominant with respect to the phase space emission of the \pkl final state at all energies
demonstrating the important role played by both N$^{*}$ and interference effects in hadron-hadron collisions.
\end{abstract}

\maketitle
\section{Introduction}
The production of strange hadrons within nuclear matter is a key ingredient in the understanding of the innermost structure 
of neutron stars (NS). Indeed, several theoretical models predict that the production of strange hadrons is energetically
favourable already at moderate densities of neutron-rich matter \cite{SchaffnerBielich:2008kb,Chatterjee:2015pua} and hence neutron stars with strange hadrons could appear.
On the other hand, the appearance of strange hadrons softens the equation of state of NS
excluding the existence of massive NS unless a strong repulsive interaction is assumed for the 
$\Lambda$NN system \cite{Lonardoni:2014bwa}. 
Since NS with two solar masses have already been measured with high precision \cite{Demorest:2010bx,Antoniadis:2013pzd},
this situation translates into a puzzle that can be solved only studying hyperons and kaons production in hadron-hadron collisions.
The best environment to carry out this kind of studies is provided by  hadron-hadron collisions at few GeV kinetic energies because
at these energies large baryonic densities, similar to those within NS, can be created. On the other hand the reaction dynamics at these
energies is dominated by hadronic resonances, that need then to be quantitatively understood \cite{Andreev:1988fj,Sarantsev:2004pe,Agakishiev:2015tfa,Fabbietti:2013npl,Agakishiev:2012xk,Agakishiev:2011qw,Agakishiev:2014nim,AbdElSamad:2010tz}.\\
For final states containing pions and nucleons produced in elementary reactions,
partial wave analysis (PWA) was already employed to correctly take into 
account interferences among resonances and determine the amplitude of the contributing waves
 \cite{Agakishiev:2016Re,Anisovich:2006bc,Anisovich:2007zz,Ermakov:2014oka}.
For the contribution of resonances to final states with open strangeness the reaction 
$\mathrm{N^*}\rightarrow$K+$\Lambda$ 
was first studied by analyzing the Dalitz plot for the reaction p+p$\rightarrow$p+K$^+$+$\Lambda$ up to kinetic energies of T = $2.5$ GeV, but without accounting for interference effects \cite{AbdElSamad:2010tz}.
The HADES collaboration was the first to employ a PWA for the search for the kaonic bound state ppK$^-$ \cite{Yamazaki200270,Yamazaki:2010mu}
in the reaction p+p$ \rightarrow$p+K$^+$+$\Lambda$ at 
a beam kinetic energy of $3.5$ GeV. 
In this reaction it was found that \nstar contribute to the measured final state and influence the background for the kaonic 
bound state \cite{Agakishiev:2014dha,Epple:2015fna}.
No evidence for the existence of ppK$^-$  bound states could be
 found and upper limits for the production of such states of the order of a few $\mu$b were extracted. 
 To get a consistent description of the open strangeness production, 
 we further improve this method and develop a framework
 that allows for the simultaneous analysis of seven different
data sets measured in the p+p$ \rightarrow$p+K$^+$+$\Lambda$ reaction
 by the COSY-TOF, HADES, DISTO and FOPI experiments in fixed target experiments
  at kinetic energies in the laboratory frame varying from 2.14 to 3.5 GeV 
\cite{Agakishiev:2014dha,Maggiora:2009gh,Maggiora:2001tx,Balestra:1999br,AbdelBary:2010pc,Muenzer:2014,Roder:2013gok}.
This is the very first joint PWA analysis of different data sets for this reaction.
 This way, the energy-dependent amplitude of seven different contributing N$^*$ resonances decaying
 into the \lamo-K$^+$ channel and for non-resonant \pkl final states could be extracted for the first time. \\
  \begin{table*}[hbt]
   \caption[Available number of events for the reaction p+p$\rightarrow$\pklo.]{List of available number of events for the reaction p+p$\rightarrow$p+K$^+$+$\Lambda$ 
  measured by the COSY-TOF, DISTO, FOPI and  HADES Collaborations. The kinetic beam energy, the total cross section and the
   reduced $\chi^2$ values resulting from different PWA analyses are shown (see text for details).}  \begin{center} 
    \begin{tabular}{cccccc}
      \hline
	\hline
      experiment  & T (GeV) &  Events/ndf & $\sigma_{tot}$ [$\mu$b]&  $\chi^{2}/ndf$(single)& $ \chi^{2}/ndf$(combined)\\
      \hline
      \DISTO \cite{Maggiora:2009gh,Maggiora:2001tx} & 2.14 & 121000 / 644 &$19.0\pm3.3$ & 0.52 & 1.52 \\
      \COSY \cite{AbdelBary:2010pc,Roder:2013gok} & 2.16 & 43662 / 712 &$19.7\pm3.5$& 1.69 & 0.44   \\
      \DISTO \cite{Maggiora:2009gh,Maggiora:2001tx}    & 2.5 & 304000 / 766 &$30.5\pm5.7$& 2.85& 2.56 \\
      \DISTO \cite{Maggiora:2009gh,Maggiora:2001tx,Balestra:1999br}   & 2.85 & 424000 / 555 &$38.7\pm7.9$& 7.68&  3.55\\
      FOPI  \cite{Muenzer:2014} & 3.1 & 903 / 226 &$43.1\pm9.3$& 1.21 &    0.91 \\
      HADES \cite{Agakishiev:2014dha}   & 3.50 & 13155 / 528 &$48.0\pm11.5$& 1.12 & 2.14 \\
      HADES \cite{Agakishiev:2014dha}   & 3.50 & 8155 / 534 &$48.0\pm11.5$& 1.38 & 1.86   \\
      \hline
	\hline
    \end{tabular}
  \end{center}
  \label{tab:samples}
\end{table*}
A second interesting aspect is the study of the p-\lam interaction.
This interaction was previously investigated primarily by means of scattering experiments \cite{SechiZorn:1969hk,Eisele:1971mk,Alexander:1969cx}.
The reaction p+p$\rightarrow$p+K$^+$+$\Lambda$ offers the possibility to study the final state interaction
of the p-\lam pair as an alternative to scattering experiments \cite{Alexander:1969cx,Roder:2013gok,Hauenstein:2016zys,Adamczewski-Musch:2016jlh}. 
Since so far the resonances were not treated in a coherent way, a precise determination of their contributions and of the scattering lengths and 
effective ranges was challenging.\\
 The combined PWA presented in this work offers the unique possibility to study the interplay between the N$^*$ coupling
 to the \lamo-K$^+$ channel and the p-\lam final state interaction.
\section{Data Samples and Combined Analysis}
%
The experimental data were measured by the COSY-TOF, DISTO, FOPI and 
HADES Collaborations. \Tab{tab:samples} provides an overview of the data sets used for the combined PWA, their beam energy and number
of events. Together with each experimental data set, simulations of the \pKL production according to phase space kinematics, 
filtered through the detector simulation and analysed as the experimental data are used for the PWA.
The details about the reconstruction of the exclusive \pKL final state, achieved resolution, efficiency, and purity 
 are explained in the already published works by the different collaborations \cite{Agakishiev:2014dha,Maggiora:2009gh,Maggiora:2001tx,Balestra:1999br,AbdelBary:2010pc,Muenzer:2014,Roder:2013gok}. The two HADES data samples at the same kinetic energy 
 correspond to two different reconstruction
 analyses including or excluding the forward spectrometer \cite{Agakishiev:2014dha}. These data sets
 are complementary and do not share any reconstructed events because of the exclusive
 selection of the final state.\\
The goal of this PWA is to employ the seven data samples in a combined analysis and extract the amplitudes of the different waves,
 characterised by their quantum numbers, leading to given final states.
We use the Bonn-Gatchina PWA (BG-PWA) framework \cite{Anisovich:2006bc,Anisovich:2007zz} 
to fit event-by-event the measured 4-momenta for the exclusive final state p+p$\rightarrow$p+K$^
+$+$\Lambda$ weighted with the coherent superposition of specific participating  waves. 
 The best choice for the waves used in the PWA is determined by comparing the experimental data to the PWA output event-by-event in terms of a log-likelihood parameter. 
 In the specific 
case of the COSY-TOF data sample, only the region of phase space within 
$ |\cos\theta^{\mathrm{CM}}_{\mathrm{p}}|<0.7$ , where $\theta^{\mathrm{CM}}_{\mathrm{p}}$ is the
proton angle in the p--p center of mass system, was considered 
because of the the poor description of the trigger 
efficiency in the simulation
for the excluded region. For the DISTO data samples the region 
corresponding to $\cos\theta^{\mathrm{CM}}_{\mathrm{p}}>0.95$ was excluded from the fit to minimize the bias introduced 
by the digitization of the scintillation-fiber sub detector used for tracking close to the target region.
These cuts were also added in the simulations used in the PWA analysis procedure.
\\
This PWA allows to decompose the baryon-baryon scattering amplitude into separate sub-processes characterized by different intermediate states. 
 \begin{table}[ht]  
 \caption[N$^*$ Resonances included in the Partial Wave Analysis.]{\nstar resonances included in the PWA written in the spectroscopic notation 
  with the corresponding masses, widths and branching ratios in the K-$\Lambda$ final states \cite{Patrignani:2016xqp,Anisovich:2011fc}.} 
  \begin{center}
    \begin{tabular}{ccccc}
      \hline
\hline
      N$^*$ & $J^{P}$ & Mass ($\frac{\mathrm{GeV}}{\mathrm{c}^{2}}$) & Width ($\frac{\mathrm{GeV}}{\mathrm{c}^{2}}$) & $\Gamma_{K\Lambda}/\Gamma_{tot}$ (\%) \\
      \hline
      1650&$\frac{1}{2}^{-}$  \cite{Patrignani:2016xqp} & 1.655 & 0.14 & 7 $\pm$ 4\\
      1710&$\frac{1}{2}^{+}$ \cite{Patrignani:2016xqp}& 1.710 & 0.23 &  15 $\pm$ 10 \\
      1720&$\frac{3}{2}^{+}$ \cite{Patrignani:2016xqp}& 1.720 & 0.25 &  4 $\pm$ 1\\
      1875&$\frac{3}{2}^{-}$ & 1.875 & 0.20 \cite{Patrignani:2016xqp} & 4 $\pm$ 2 \cite{Anisovich:2011fc} \\
      1880&$\frac{1}{2}^{+}$ & 1.870 & 0.24 \cite{Anisovich:2011fc} & 2 $\pm$ 1\cite{Anisovich:2011fc} \\
      1895&$\frac{1}{2}^{-}$ & 1.895 & 0.09 \cite{Anisovich:2011fc} &18 $\pm$ 5 \cite{Anisovich:2011fc} \\
      1900&$\frac{3}{2}^{+}$ & 1.900 & 0.26  \cite{Anisovich:2011fc}& 11 $\pm$ 9 \cite{Patrignani:2016xqp}\\
      \hline
\hline
    \end{tabular}
  \end{center}

  \label{tab:bg-pwa:nstars} 
\end{table}
Within the BG-PWA framework this is achieved by fitting event-by-event the experimental 4-vectors for a given reaction measured within the acceptance
of the spectrometer with a coherent superposition of the participating waves. This coherent cocktail of contributing waves is weighted with the full scale 
phase space simulations of the considered final state that accounts for the geometrical acceptance and reconstruction efficiency of the spectrometer. \\
 Within the BG-PWA, the production cross section of a three particle final state with single particle four-momenta $q_{1,2,3}$ is parametrized as 
 \cite{Anisovich:2006bc}:
\begin{equation}
  d\sigma = \frac{(2\pi)^{4} |A|^{2}}{4|\vec{k}|\sqrt{s}} \text{d}\Phi_{3}\left(P,q_{1},q_{2},q_{3}\right),
  \label{equ:bg-pwa:crosssection}
\end{equation}
wherein $P$ is the total four-vector, $\vec{\mathrm{k}}$ is the beam momentum, $\sqrt{s}$ the center of mass energy of the reaction, $d\Phi_{3}$ is the infinitesimal phase-space volume of the final state and $A$ is the total transition amplitude of the considered reaction. Both initial and  final states can be seen as a superposition of eigenstates with various angular momentum and $A$ is the sum over all the transition amplitudes $A^{\alpha}_{tr}$ between these eigenstates \cite{Ermakov:2011at}:
\begin{equation}
  \begin{split}
    A = \sum_{\alpha} A^{\alpha}_{\text{tr}}(s) Q_{\mu_{1}..\mu_{j}}^{\text{in}}(S,L,J) A_{2b}\left(i,S_{2},L_{2},J_{2}\right) \\
    Q_{\mu_{1}..\mu_{j}}^{\text{fin}}(i,S_{2},L_{2},J_{2},S^{\prime},L^{\prime},J).
  \end{split}
  \label{equ:bg-pwa:totalamplitude}
\end{equation}
The index $\alpha$ runs over all the amplitudes contributing to the transition from the initial to the final state.
The factors $Q_{\mu_{1}..\mu_{j}}^{\text{in}}(S,L,J)$ and $Q_{\mu_{1}..\mu_{j}}^{\text{fin}}(i,S_{2},L_{2},J_{2},S^{\prime},L^{\prime},J)$ are the spin-momentum
operators of the initial and final states respectively and the indexes $\mu_{j}$ refer to the rank of the total angular momentum 
$J$ in the spin--momentum operators $Q$. The index $i$ refers to the two-particle sub-system considered in the final state. 

The dependency of the amplitudes $A^{\alpha}_{\text{tr}}(s)$ upon the centre of mass energy is given by:
\begin{equation}
  A^{\alpha}_{\text{tr}}(s) = \left(a^{\alpha}_{1} + a^{\alpha}_{3}\sqrt{s}\right) exp\left(\text{i}a_{2}^{\alpha}\right).
  \label{equ:bg-pwa:partialamplitude}
\end{equation}
The real parameters $a_{1}^{\alpha}$,  $a_{2}^{\alpha}$ and $a_{3}^{\alpha}$ are determined by the fit 
to the experimental data. \\
The parametrization of the factor $A_{2b}$ depends on the final state.
For the production of a \nstar resonance, the final state is treated as a two-body system composed of a proton and the \nstaro.
In this case the quantum numbers $S_2,\,L_2,\,J_2$ refers to the \nstaro, while the $S^{\prime},L^{\prime},J$ represent 
the quantum numbers of the 
\nstaro-proton system.
Non resonant \pkL final states are also treated as a two particle system composed of a p\lam "particle" and a \Kpo.
In this case $S_2,\,L_2,\,J_2$ are the spin, angular and total angular momentum of the p\lam 'particle' while $S^{\prime},L^{\prime},J$ are the 
quantum numbers of the p\lamo-K$^+$ system.\\
For the resonant case, the factor $A_{2b}$ is parametrized with a relativistic Breit-Wigner formula \cite{Jackson:1964zd}.
\begin{equation}
  A_{2b}^{\beta} = \frac{1}{(M^{2}-s-\text{i}\Gamma M)},
  \label{equ:bg-pwa:breitwigner}
\end{equation}
with $M$ and $\Gamma$ as the pole mass and width of the corresponding resonance. 
For the presented analysis, the
\nstar resonances listed in \tab{tab:bg-pwa:nstars} have been considered with fixed masses and fixed widths taken from \cite{Patrignani:2016xqp,Anisovich:2011fc}. \\
 To obtain an acceptable description of the experimental data it is necessary 
 to include non-resonant partial wave amplitudes. We have 
 included these amplitudes in a simple form which provides a correct behaviour
near threshold. For the S-wave this form corresponds to the well known Watson-Migdal parameterization.
The resulting $A_{2b}$ amplitude is
\begin{equation}
  A_{2b}^{\beta} = \frac{\sqrt{s_{i}}}{1-\frac{1}{2}r^{\beta}q^{2}a^{\beta}_{p\Lambda} + \text{i}qa^{\beta}_{p\Lambda} q^{2L}/F\left(q,r^{\beta},L\right)},
\label{equ:bg-pwa:plscattering}
\end{equation}
where q is the p-\lam relative momentum, $a_{p-\Lambda}^{\beta}$ is the p-\lam-scattering length,
$r^{\beta}$ is the effective range of the p-\lam system and the index $\beta$ denotes the quantum
numbers combination.\\
\begin{figure*}
   \begin{picture}(100,30)
      \put(0,0){\makebox(100,30)[t]{\includegraphics[width=100\unitlength,height=30\unitlength]{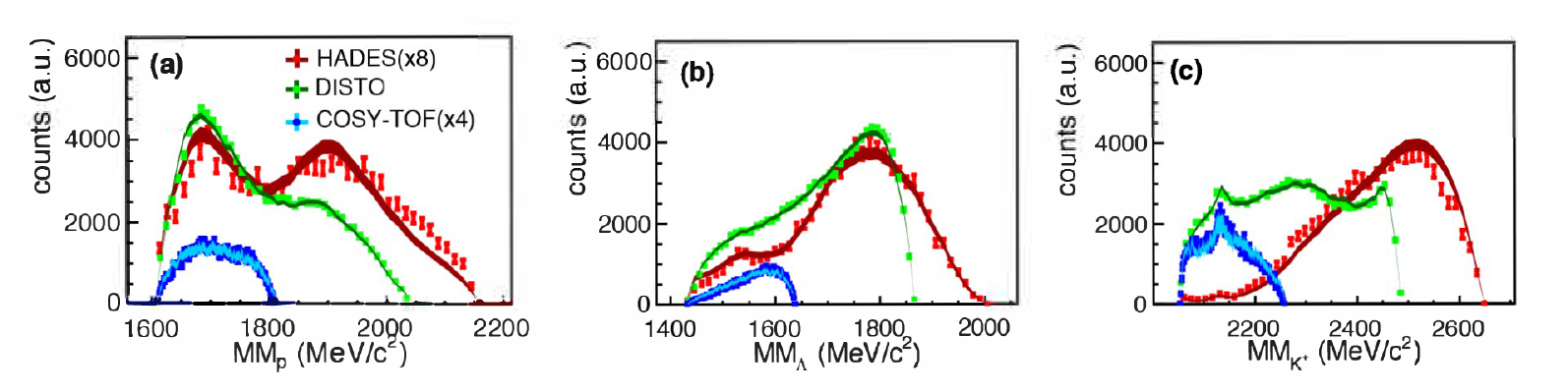}}}
      \end{picture} 
 \caption{(Color online). Missing mass distributions (MM) for the three different particles of the final state (p, $\Lambda$, K$^+$) are shown. 
 The experimental data within the geometrical acceptance are from COSY-TOF at $2.16$ GeV (blue symbols), DISTO at $2.85$ GeV (green symbols) and HADES at $3.5$
GeV (red symbols) samples. The colored lines in the same color-code represent the PWA results (see text for details).}   
 \label{fig:results:combined}
\end{figure*}
\begin{figure*}
  \begin{picture}(100,3)
  \end{picture}
  \begin{picture}(100,26.2)
     \put(0,0){\makebox(33.3,26.2)[t]{\includegraphics[width=33.3\unitlength,height=26.2\unitlength]{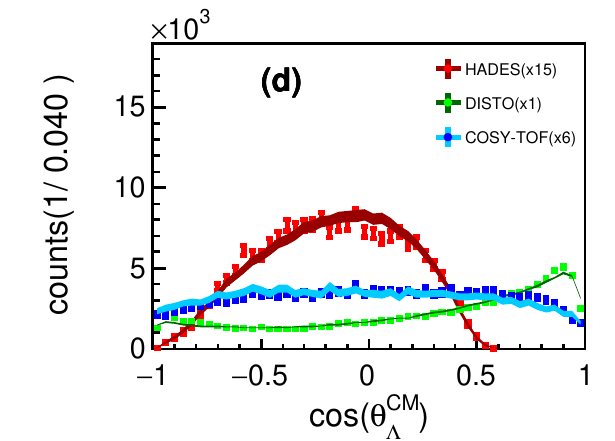}}}
     \put(33.3,0){\makebox(33.3,26.2)[t]{\includegraphics[width=33.3\unitlength,height=26.2\unitlength]{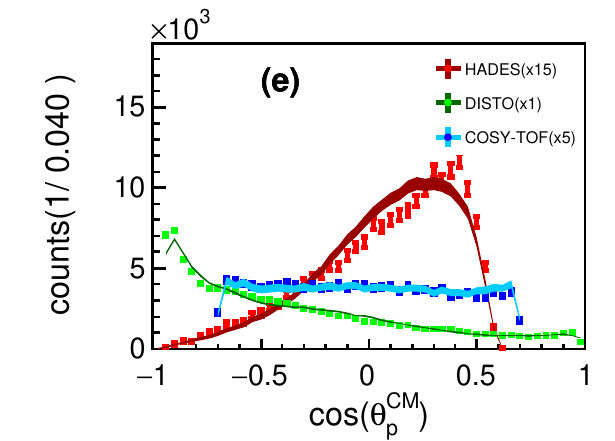}}}
     \put(66.6,0){\makebox(33.3,26.2)[t]{\includegraphics[width=33.3\unitlength,height=26.2\unitlength]{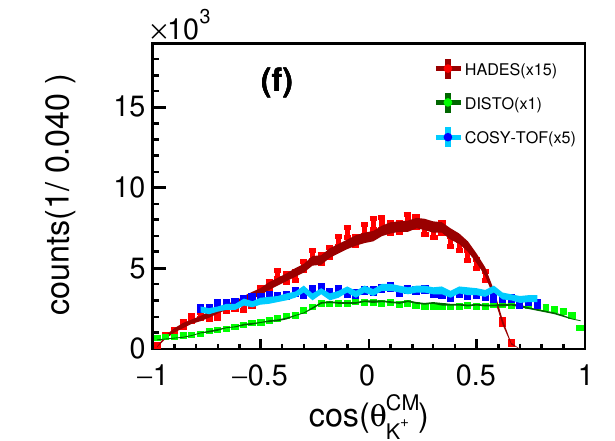}}}
     \put(16.66,-3){(d)}
     \put(50,-3){(e)}
     \put(83.33,-3){(f)}
  \end{picture}
  \begin{picture}(100,3)
  \end{picture}
  \begin{picture}(100,26.2)
    \put(0,0){\makebox(33.3,26.2)[t]{\includegraphics[width=33.3\unitlength,height=26.2\unitlength]{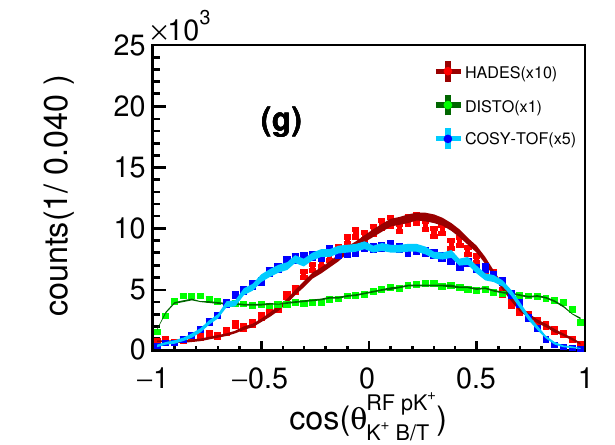}}}
    \put(33.3,0){\makebox(33.3,26.2)[t]{\includegraphics[width=33.3\unitlength,height=26.2\unitlength]{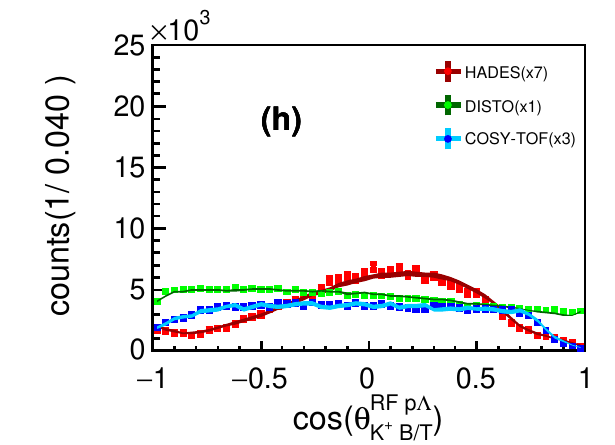}}}
    \put(66.6,0){\makebox(33.3,26.2)[t]{\includegraphics[width=33.3\unitlength,height=26.2\unitlength]{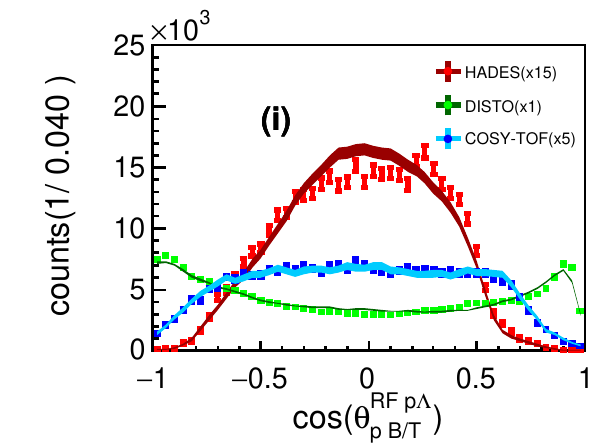}}}
    \put(16.66,-3){(g)}
    \put(50,-3){(h)}
    \put(83.33,-3){(i)}
  \end{picture}
  \begin{picture}(100,3)
  \end{picture}
  \begin{picture}(100,26.2)
    \put(0,0){\makebox(33.3,26.2)[t]{\includegraphics[width=33.3\unitlength,height=26.2\unitlength]{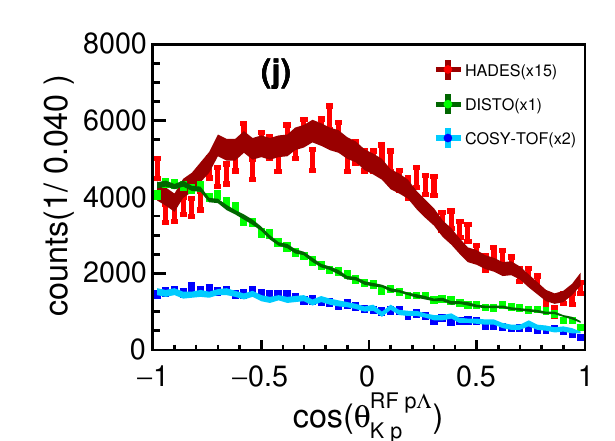}}}
    \put(33.3,0){\makebox(33.3,26.2)[t]{\includegraphics[width=33.3\unitlength,height=26.2\unitlength]{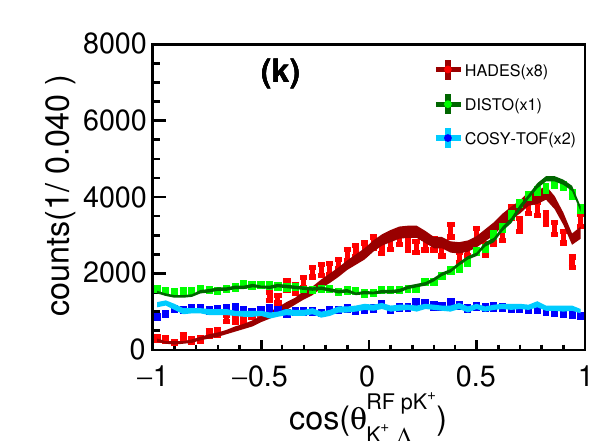}}}
    \put(66.6,0){\makebox(33.3,26.2)[t]{\includegraphics[width=33.3\unitlength,height=26.2\unitlength]{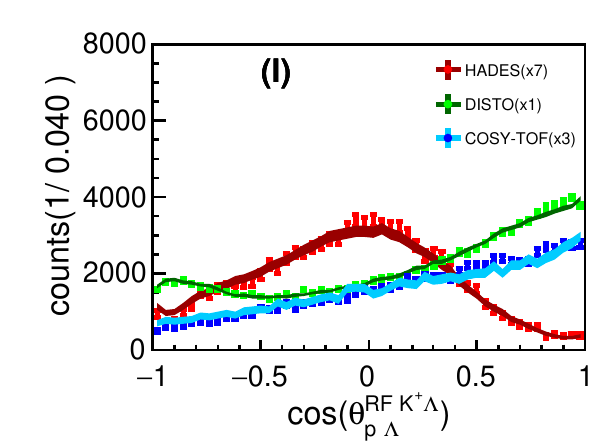}}}
    \put(16.66,-3){(j)}
    \put(50,-3){(k)}
    \put(83.33,-3){(l)}
    \end{picture}
  \begin{picture}(100,3)
  \end{picture}
  \caption{
  Angular correlations for the \pkl final state.
   The upper index at the angle indicates
the rest frame (RF) in which the angle is displayed. The lower index
names the two particles between which the angle is evaluated. CM
stands for the center-of-mass system. B and T denote the beam
and target vectors, respectively. The observables are: 
  CM distributions ($cos\left(\theta^{CM}_{X}\right) $) of the \lam (d), Proton (e) and Kaon (f);
Gottfried-Jackson distributions $\cos\left(\theta_{K B/T}^{RF\, pK}\right)$ (g),$\cos\left(\theta_{K B/T}^{RF\, K\Lambda}\right)$ (h),
    $\cos\left(\theta_{p B/T}^{RF\, p\Lambda}\right)$ (i) and Helicity angle distributions 
    $\cos\left(\theta_{Kp}^{RF \, p\Lambda}\right)$ (j),$\cos\left(\theta_{K\Lambda}^{RF \, pK}\right)$ (k)and $\cos\left(\theta_{p\Lambda}^{RF \, K\Lambda}\right)(l)$.
     The experimental data within the geometrical acceptance are from COSY-TOF at $2.16$ GeV (blue symbols), DISTO at $2.85$ GeV (green symbols) and HADES at $3.5$
GeV (red symbols) samples. The colored lines in the same color-code represent the PWA results.
  }
  \label{fig:results:angles}
\end{figure*}
$F(q,r,L)$ is the Blatt-Weisskopf factor used for the normalization, it is 1 for L=0 and 
the explicit form for other partial waves can be found in \cite{Anisovich:2006bc}.
 The values of the 
scattering length and effective range can be set as free parameters in the PWA fit and hence be extracted within this analysis. 
This coherent approach differs from the analysis techniques usually employed for the extraction of scattering parameters
\cite{Gasparyan:2003cc} and should be considered as complementary.
\\
Another intermediate channel contributing to \pkL final state is the \sigo-N cusp, which appears at or 
above the $\Sigma$-N threshold (2130 MeV/$c^{2}$) \cite{ElSamad:2012kg}. 
The coupling between the \sigo-N and \lamo-N
channels  leads to an enhancement of the cross-section in the p-\lam final state in a mass range close to the
above mentioned threshold.
In order to include the cusp contribution in the BG-PWA framework, new transition waves must be added to Eq. \ref{equ:bg-pwa:totalamplitude}. Since the cusp is 
located at the 
\sigo-N threshold, the $\Sigma$ and N must be in a relative S-wave state, which means that the spin-parity of the \sigo-N system is either $J^{P} =0^{+}$ or $1^{+}$ 
\cite{ElSamad:2012kg}. The resulting p-\lam system then may appear in an S-wave state in case of $J^{P}=0^{+}$ or in an s- or d-wave state in case of $J^{P}=1^{+}$. 
This has also been confirmed by an analysis of the \sigo-N cusp carried out by the COSY-TOF collaboration \cite{ElSamad:2012kg}. 
Additionally, since the cusp is a resonance structure in analogy to the \nstar, the Breit-Wigner 
parametrization is used for A$_{2b}$ (Eq. \ref{equ:bg-pwa:breitwigner}) where the mass and width 
are varied within $2.1-2.16$ GeV and $0.01-0.03$ GeV/c$^2$, respectively in the PWA fit. 
This first attempt can be also replaced by a more sophisticated parametrization of the cusp contribution like
a Flatte' function, but this is beyond the scope of this investigation. Indeed the cusp contribution has a negligible effect on 
the determination of the \nstaro contributions.\\

\section{Results}
\label{sect:results:method}
First, the PWA was performed individually for the different data samples to determine the correct start values of 
the parameters for the global fit.
The total number of available degrees of freedom for each data set is listed in \tab{tab:samples}. 
The total number of free parameters in the PWA fit containing all accessible N$^{*}$ is equal to $345\pm17$, the
error refers to the systematic variation of the contributing  N$^{*}$ considered in the global fit.
The best solution of the PWA fit corresponds to the minimum of the log-likelihood obtained by fitting the experimental data
with the PWA event-by-event.\\
A comparison of the three missing mass spectra and CM,
Gottfried-Jackson and Helicity angle distributions (for the definition of these variables see \cite{Agakishiev:2011qw})
obtained from the experimental data and from the single PWA fits was carried out and the corresponding reduced $\chi^{2}$
values are listed in \tab{tab:samples}. Only the statistical error of the experimental data has been considered to
evaluate the $\chi^{2}$ of the single PWA fits. 
As a second step, a simultaneous PWA of three data samples was carried out. This
intermediate step allowed to determine the starting values for the global fit. 
The HADES, FOPI and DISTO (T $=\, 2.5$ GeV) samples were selected to account for both the contributions
from the $\Sigma$-N cusp and from higher mass resonances.  
 After finding a solution that described the three data samples, further data samples were 
 added stepwise.
The starting values of each new PWA fit were taken from the results of the previous fit step.
The systematic error of the experimental samples have not been considered in the fit since
the latter were not available for all the data sets.
To account for possible systematic variations of the 
kinematic distributions we have considered all permutations for the exclusion of one or more
 \nstar resonances from the list in \tab{tab:bg-pwa:nstars} in the PWA fit. The five best solutions 
 in terms of log-likelihood 
 obtained from this systematic variation of the PWA
  fits were considered to extract the final results and the PWA systematic errors.
As far as the resonances are concerned, considering the list of seven resonances in \tab{tab:bg-pwa:nstars},
  the five best solutions  correspond to the following combinations: 1) all seven \nstar included, 2) N$^{*}$(1720) excluded,
  3) N$^{*}$(1875) excluded,  4) N$^{*}$(1900) excluded and,
  5) N$^{*}$(1900) and  N$^{*}$(1875) excluded.  \\
The reduced $\chi^{2}$ values for the combined PWA listed in \tab{tab:samples} were obtained by comparing the experimental 
data in the mass and angle variables with the average values of the five best PWA solutions, taking as errors the
statistical errors of the experimental data and the standard deviation of the five solutions for each bin.  By adding additional solutions
the $\chi^{2}$  did not improve. This justifies the choice of the five best solutions.
A more refined treatment of systematic uncertainties is current under development.
\\
 \begin{figure*}[ht]
   \begin{picture}(100,30.)
     \put(0,0){\makebox(100,30.)[t]{\includegraphics[width=100\unitlength,height=35.\unitlength]{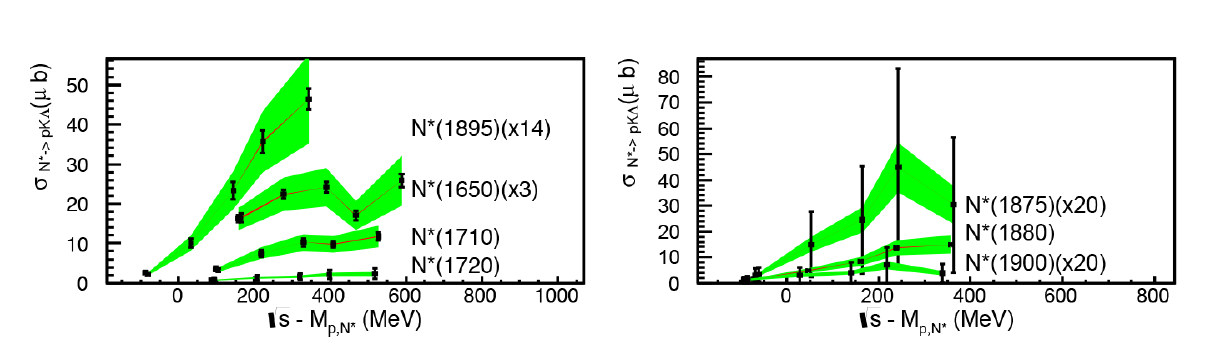} }}
  \end{picture} 
  \begin{picture}(100,3)
  \end{picture}
  \caption{(Color online). Cross sections of the different \nstar resonances decaying into the \pKL final state 
  obtained from the combined PWA as a function
  of the excess energy. The excess energy is calculated as the center of mass energy
of the p--p colliding system minus the sum of the proton $\Lambda$ and Kaon masses ($\sqrt{s} - \mathrm{M}_{p,K^+,\Lambda}$)
  The black bars show the systematic errors originating from the five different PWA solutions and the green bands
  represent the errors due to the normalization to the total \pkl cross section.}
  \label{fig:results:excitation:resonant}
\end{figure*}
\begin{figure*}[ht]
  \begin{picture}(100,30.)
    \put(0,0){\makebox(100,30.)[t]{\includegraphics[width=100\unitlength,height=35.\unitlength]{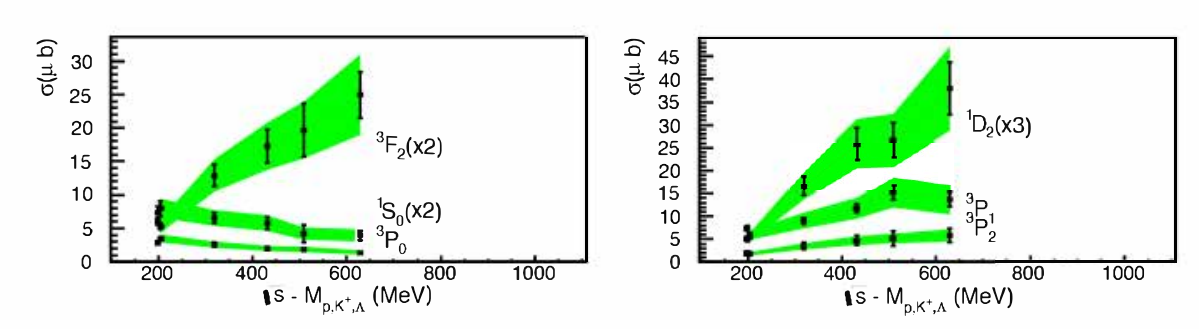} }}
  \end{picture} 
  \begin{picture}(100,3)
  \end{picture}
  \caption{(Color online). Cross sections of the initial state waves as a function of the excess energy 
  for the \pkL final state. The excess energy is calculated as the center of mass energy
of the p--p colliding system minus the sum of the proton and N* masses ($\sqrt{s} - \mathrm{M}_{p,N^*}$).
  The error bars correspond to the standard deviation among the five best PWA solutions and the green 
  band refers to the normalization
  to the total \pKL production cross section.}
  \label{fig:results:excitation:initial}
\end{figure*}
\fig{fig:results:combined} shows the missing mass distributions (MM) for the three final state particles p, \lam and K$^+$ for
COSY-TOF at $2.16$ GeV (blue symbols), DISTO at $2.85$ GeV (green symbols) and HADES at $3.5$
GeV (red symbols) data samples measured within their respective acceptances and arbitrarily normalized. 
The signature of the $\Sigma$-N cusp is visible in the COSY-TOF and DISTO MM$_{\mathrm{K^+}}$ distribution around $2.13$ GeV/$c^2$.
 The errors of the experimental data are statistical only. The lines in the same color-code represent the PWA 
 results for the corresponding data sets.
The line widths represent the error bands of the global PWA fit expressed as the standard deviation of
the five best PWA solutions. 
\fig{fig:results:angles} shows the angular distributions of the three particles measured in the final state for different reference systems
for the same data samples discussed in \fig{fig:results:combined}.
A similar quality is obtained for the description of the kinematic variables of other data samples.\\
 The output of each PWA solution provides the strength of the individual waves
with respect to the total measured yield. 
The resulting relative contributions 
of the resonant and non-resonant waves 
  can be translated into cross sections for the K$\Lambda$ decay channel multiplying the relative yield by the 
total production cross section  for the \pkl final state.\\
The total \pkl cross section  for the different data sets was evaluated employing a phase space fit of the
existing measurements of the \pkl channel as a function of the excess energies 
\cite{AbdElSamad:2010tz,AbdelBary:2010pc,ElSamad:2012kg,AbdelSamad:2006qu,id1122}.
The error associated to the \pkl cross section of each data sample is extracted from the fit. 
A detailed description of the extraction of the \pkl cross sections can be found in \cite{MuenzerHyp}. \\
In \Fig{fig:results:excitation:resonant} the cross section for the different \nstar channels decaying into
the K$\Lambda$ final state is plotted versus its excess energy calculated as the center of mass energy
of the p--p colliding system minus the sum of the proton and N* masses ($\sqrt{s} - \mathrm{M}_{p,N^*}$).
The standard deviation of the five best solutions is shown by the black vertical error bars, the green bands show the error originating
 from the cross section normalization. 
The non-vanishing cross section below the respective thresholds is due to the large width of all the considered resonances (see Table \ref{tab:bg-pwa:nstars}).
The relative contribution of the non-resonant amplitude decreases from 37\% for 2.14 GeV to 10\% for 3.5 GeV, so
that most of the yield stems from N$^{*}$ resonances for all the measured energies.
The dominant contribution from the \nstar resonances is consistent with the results shown in Ref. 
\cite{AbdElSamad:2010tz}, except for the relative
contribution of the N$^*(1650)$, which is decreasing as a function of the beam energy in \cite{AbdElSamad:2010tz}.
In this work we found an increment of the N$^*(1650)$
 similarly to the N$^*(1710)$ and N$^*(1720) $. This difference probably results from neglecting
interference in Ref. \cite{AbdElSamad:2010tz}.\\
The \sigo-N cusp contribution varies from $10^{-3}$ to $10^{-2}$ with decreasing energy with respect to the N$^*$ and is not shown in \fig{fig:results:excitation:resonant}.
The global PWA fit favors the \sigo-N cusp contribution of the s- or d-wave state $J^{P}=1^{+}$ 
 with respect to the S-wave $J^{P}=0^{+}$ as shown by the amplitudes in Table \ref{NsCS2}. 
The obtained \sigo-N cusp yield  is slightly different from the findings in Ref. \cite{ElSamad:2012kg} where at a beam energy of $2.28$ GeV the contribution
of the cusp was found equal to $5$\% of the total cross section, but neglecting interferences. \\
 Figure \ref{fig:results:excitation:initial} shows the 
cross sections of the different p+p initial states as a function of the \pkL
 excess energy calculated as the center of mass energy
of the p--p colliding system minus the sum of the proton $\Lambda$ and Kaon masses ($\sqrt{s} - \mathrm{M}_{p,K^+,\Lambda}$).
  The error bars are associated to the standard deviation of the five best PWA solutions, and the green 
  band refers to the uncertainty of the exclusive \pKL production cross section. 
  All extracted cross-sections as a function of the excess energy are summarised in \Tab{NsCS2} and \Tab{IS}. \\
The non-resonant amplitude included in this PWA is parametrized as a function of the scattering
length and effective range for the p-\lam final state interaction. 
The interference of the non-resonant partial waves with the resonant amplitudes allows us to extract independently the values
 for S-wave singlet and S-wave triplet partial waves.
  In \Tab{tab:results:excitation:scatteringlength} the resulting values for the scattering lengths are listed. The values 
 are obtained by averaging the five best PWA solutions. The first error represents the standard deviation of the five fit results. 
 The second one is the PWA fit error obtained by adding quadratically the 
PWA fit errors from the five solutions.  In the same table also 
the scattering lengths obtained from p+p reactions with unpolarized \cite{Alexander:1969cx,Budzanowski:2010ib} and 
 polarized beams \cite{Hauenstein:2016zys} and the predictions by recent theoretical calculations \cite{Haidenbauer:2013oca,ESC08} are shown. \\
 The results from this PWA are comparable with previously extracted values.
 Different parametrization, as by means of a Jost function,
 might modify the extracted scattering parameters. Still, the comparison of the values that have been extracted
 within this PWA to other experimental results and theoretical parametrisation demonstrate that despite of the very large
 number of free parameters of this PWA and that all contributions have been treated coherently, a reasonable agreement is achieved.
 

\begin{table*}
 \caption{Scattering lengths extracted from the combined PWA fit and reference values from previous measurements \cite{Alexander:1969cx,Budzanowski:2010ib,Hauenstein:2016zys} and theoretical calculations \cite{Haidenbauer:2013oca,ESC08} (see text for details).}
  \label{tab:results:excitation:scatteringlength}  \begin{center}
    \begin{tabular}{ccc}
	\hline
      \hline
  Source &  ${}^{1}S_{0}$ $a_{\Lambda-p}$ [fm] & ${}^{3}S_{1}$ $a_{\Lambda-p}$ [fm]   \\
\hline
     This work & $   -1.43\pm 0.36 \pm 0.09 $  &   $-1.88\pm 0.38 \pm 0.10 $ \\
          \cite{Alexander:1969cx} & $-1.8 ^{+2.3}_{-4.2}$  & $-1.6 ^{+1.1}_{-0.8}$\\ 
          \cite{Budzanowski:2010ib} & $-2.43^{+0.16}_{-0.25}$  & $-1.56^{0.19}_{-0.22}$    \\ 
           \cite{Hauenstein:2016zys} &- &$-2.55^{+0.72}_{-1.39}\pm 0.6 \pm 0.3$  \\
           
      \hline
      
      $\chi$EFT LO \cite{Haidenbauer:2013oca} &$-1.91$  & $-1.23$ \\
       $\chi$EFT NLO \cite{Haidenbauer:2013oca} & $-2.91$ & $-1.54$  \\
        ESC08 \cite{ESC08} &$-2.7$ & $-1.65$ \\
	\hline
	\hline
          \end{tabular}
  \end{center}
\end{table*}
\section{Summary}
We have applied a combined PWA to seven different data sets measuring the reaction p+p$\rightarrow$p+K$^+$+$\Lambda$
for kinetic energies between $2.14$ and $3.5$ GeV and determined for the first time the production amplitude of the resonances: N$^{*}(1650)1/2^{-}$, N$^{*}(1710)1/2^{+}$,  
N$^{*}(1720)3/2^{+}$, N$^{*}(1875)3/2^{-}$, N$^{*}(1880)1/2^{+}$, N$^{*}(1895)1/2^{-}$ and N$^{*}(1900)3/2^{+}$ and initial state partial wave
as a function of the excess energy. The contribution of the resonances has been found to be dominant with respect to the 
direct production of the \pkl final state especially for the highest kinetic energy of 3.5 GeV where 90\% of the yield is associated to N*.
This shows not only that the resonant production is dominating this energy regime of hadron-hadron collisions, but also provides
a quantitative understanding for the first time of the interference effects on the N* excitation function. The \sigo-N cusp was also included in the PWA but its contribution is found
to vary between $10^{-3}$ to $10^{-2}$ with decreasing energy. Hence it does not influence the obtained results for the N* and non resonant amplitudes.
The p-\lam scattering lengths have also be extracted from this combined PWA and found to be consistent with previous 
measurements. Higher precision should be achieved with a dedicated analysis of the data at the lowest energies of the here presented data samples.
A natural improvement of the results presented in this work will be achieved by including two additional
data sets measured by the COSY-TOF collaboration at $2.7$ and $2.95$ GeV \cite{Hauenstein:2016zys,Jowzaee:2015qda}.
\section{Acknowledgements}
The authors acknowledge the support by funding the following funding agencies: DFG, Grant FA 898/2-1 and NCN 2016/23/P/ST2/04066 POLONEZ.
\begin{table*}
\caption[]{Production cross sections of the total \pKL non-resonant contribution and of the different N$^{*}$ resonances decaying into the \pKL
final state obtained from the global PWA as a function of the beam kinetic energy. 
The cross sections refer to the amplitudes prior to the coherent sum of the latter and hence do not consider interference effects. The N$^*$ cross sections 
are not corrected for the branching ratio into the K$^+$-$\Lambda$ final states. The first error corresponds to the systematic error due to the five best 
solutions, the second stems from the cross section normalisations. The systematic error of the PWA fitting procedure is found to be negligible and hence is 
not shown.  }
\begin{tabular}{cccc}
\hline
\hline
& 3.500 GeV & 3.100 GeV &2.85 GeV\\ 
\hline
pK$^{+}\Lambda$ [$\mu$b]   &$      5.1\pm 1.0\pm 1.2$&$      6.3\pm 1.2\pm 1.4$&$      6.5\pm 1.1\pm 1.3$ \\ 
 N$^{*}(1650)\rightarrow\,$\pKL [$\mu$b]      &$       8.6\pm 0.6\pm 2.1$&$      5.7\pm 0.4\pm 1.2$&$      8.1\pm 0.4\pm 1.6$\\
 N$^{*}(1710)\rightarrow\,$\pKL [$\mu$b]      &$     11.7\pm 1.0\pm 2.8$&$      9.7\pm 0.8\pm 2.1$&$     10.2\pm 1.0\pm 2.1$\\
 N$^{*}(1720)\rightarrow\,$\pKL [$\mu$b]      &$    2.4\pm 1.3\pm 0.6$&$      2.2\pm 1.2\pm 0.5$&$      1.6\pm 0.8\pm 0.3$\\ 
N$^{*}(1875)\rightarrow\,$\pKL [$\mu$b]      &$     1.5\pm 1.3\pm 0.4$&$      2.2\pm 1.9\pm 0.5$&$      1.2\pm 1.0\pm 0.2$\\
 N$^{*}(1880)\rightarrow\,$\pKL [$\mu$b]      &$     14.9\pm 0.2\pm 3.6$&$     13.7\pm 0.4\pm 3.0$&$      8.4\pm 0.4\pm 1.7$ \\
 N$^{*}(1895)\rightarrow\,$\pKL [$\mu$b]      &$      3.3\pm 0.2\pm 0.8$&$      2.5\pm 0.2\pm 0.6$&$      1.7\pm 0.2\pm 0.3$\\
 N$^{*}(1900)\rightarrow\,$\pKL [$\mu$b]      &$      0.2\pm 0.2\pm 0.0$&$      0.3\pm 0.3\pm 0.1$&$      0.2\pm 0.2\pm 0.0$\\
$\Sigma-$N$(1^{+}$S) [$\mu$b] &$      0.01\pm 0.02\pm 0.002 $&$      0.03\pm 0.02\pm 0.007 $&$      0.12\pm 0.05\pm 0.02$\\
$\Sigma-$N$(1^{+}$D) [$\mu$b] &$      0.13\pm 0.02\pm 0.03 $&$      0.2\pm 0.04\pm 0.05 $&$      0.5\pm 0.08\pm 0.1$\\ 
\hline
\end{tabular}
\begin{tabular}{cccc}
\hline
\hline
& 2.5 GeV & 2.157 GeV &2.14 GeV\\ 
\hline
pK$^{+}\Lambda$ [$\mu$b]& $      7.2\pm 1.1\pm 1.3$&$      7.5\pm 0.6\pm 1.3$&$      7.1\pm 0.6\pm 1.2$   \\ 
N$^{*}(1650)\rightarrow\,$\pKL [$\mu$b]  &$      7.5\pm 0.4\pm 1.4$&$      5.5\pm 0.3\pm 1.0$&$      5.4\pm 0.3\pm 1.0$      \\ 
N$^{*}(1710)\rightarrow\,$\pKL [$\mu$b]  & $      7.5\pm 0.9\pm 1.4$&$      3.3\pm 0.5\pm 0.6$&$      3.5\pm 0.5\pm 0.6$    \\ 
 N$^{*}(1720)\rightarrow\,$\pKL [$\mu$b]  & $      1.3\pm 0.7\pm 0.2$&$      0.8\pm 0.4\pm 0.1$&$      0.7\pm 0.3\pm 0.1$   \\ 
 N$^{*}(1875)\rightarrow\,$\pKL [$\mu$b]  &$      0.7\pm 0.6\pm 0.1$&$      0.2\pm 0.1\pm 0.0$&$      0.2\pm 0.1\pm 0.0$    \\ 
N$^{*}(1880)\rightarrow\,$\pKL [$\mu$b]  &$      4.8\pm 0.3\pm 0.9$&$      1.7\pm 0.1\pm 0.3$&$      1.5\pm 0.1\pm 0.3$     \\ 
 N$^{*}(1895)\rightarrow\,$\pKL [$\mu$b]  &$      0.7\pm 0.1\pm 0.1$&$      0.2\pm 0.0\pm 0.0$&$      0.2\pm 0.0\pm 0.0$    \\ 
N$^{*}(1900)\rightarrow\,$\pKL [$\mu$b]  & $      0.2\pm 0.2\pm 0.0$&$      0.1\pm 0.1\pm 0.0$&$      0.1\pm 0.1\pm 0.0$     \\ 
$\Sigma-$N$(1^{+}$S) [$\mu$b]&$      0.12\pm 0.04\pm 0.02$&$      0.16\pm 0.04\pm 0.03$&$      0.13\pm 0.03\pm 0.02$\\
$\Sigma-$N$(1^{+}$D) [$\mu$b]&$      0.34\pm 0.07\pm 0.06$&$      0.21\pm 0.04\pm 0.04$&$      0.17\pm 0.03\pm 0.03$\\
\hline
\hline
\end{tabular}
\label{NsCS2}
\end{table*}

\begin{table*}
\caption[]{Contributions of the different initial state waves as a function of the beam kinetic energy. 
The obtained cross sections are normalised to the exclusive \pKL cross section. The first error corresponds to the systematic error due to the five best solutions, the second originates from the cross section normalisation. The systematic error of the PWA fitting procedure is found to be negligible and hence is not shown.}
\begin{tabular}{ccccccc}
\hline
\hline
& 3.5 GeV & 3.1 GeV& 2.85 GeV & 2.5 GeV& 2.157 GeV & 2.140 GeV \\ 
\hline
$\sigma_{\mathrm{pk}\Lambda}$ [$\mu$b] & 48.0$\pm$5.8 & 43.1$\pm$5.3& 38.7$\pm$4.8 & 30.5$\pm$3.9 & 19.7$\pm$2.7 & 19.0$\pm$2.6  \\
 $     {}^{1}$S$_{0}$[$\mu$b]&$       2.0\pm 0.4\pm 0.5$&$      2.1\pm 0.6\pm 0.5$&$     2.9\pm 0.5\pm 0.6$&$      3.3\pm 0.4\pm 0.6$&$     4.0\pm 0.6\pm 0.7$&$      3.7\pm 0.5\pm 0.6$\\ 
 $     {}^{1}$D$_{2}$[$\mu$b]&$     12.7\pm 1.9\pm 3.0$&$      8.9\pm 1.2\pm 1.9$&$      8.6\pm 1.2\pm 1.8$&$      5.6\pm 0.7\pm 1.0$&$     1.8\pm 0.1\pm 0.3$&$      2.5\pm 0.2\pm 0.4$\\ 
 $     {}^{3}$P$_{0}$[$\mu$b]&$       1.4\pm 0.2\pm 0.3$&$      1.8\pm 0.2\pm 0.4$&$      2.1\pm 0.3\pm 0.4$&$      2.6\pm 0.3\pm 0.5$&$      3.5\pm 0.3\pm 0.6$&$      2.9\pm 0.2\pm 0.5$\\ 
 $     {}^{3}$P$_{1}$[$\mu$b]&$     13.7\pm 1.4\pm 3.3$&$     15.3\pm 1.6\pm 3.3$&$    11.8\pm 0.9\pm 2.4$&$     9.1\pm 0.7\pm 1.7$&$      5.9\pm 0.6\pm 1.0$&$      5.1\pm 0.5\pm 0.9$\\ 
 $     {}^{3}$P$_{2}$[$\mu$b]&$       5.8\pm 1.5\pm 1.4$&$      5.2\pm 1.5\pm 1.1$&$      4.8\pm 1.1\pm 1.0$&$      3.5\pm 0.8\pm 0.7$&$      1.8\pm 0.4\pm 0.3$&$      1.9\pm 0.4\pm 0.3$\\ 
 $     {}^{3}$F$_{2}$[$\mu$b]&$     12.5\pm 1.7\pm 3.0$&$      9.8\pm 2.0\pm 2.1$&$      8.6\pm 1.2\pm 1.8$&$      6.4\pm 0.8\pm 1.2$&$      2.7\pm 0.3\pm 0.5$&$      3.0\pm 0.3\pm 0.5$\\ 
\hline \hline 
\end{tabular}
\label{IS}
\end{table*}
\bibliography{pwa_paper_main}

\end{document}